\documentclass[twocolumn,aps,pra,amsmath,amssymb,showpacs,superscriptaddress]{revtex4}

\usepackage{ucs}
\usepackage[utf8]{inputenc} 

\usepackage{graphicx}
\usepackage{graphics}
\usepackage{latexsym}
\usepackage{amsmath}
\usepackage{amsfonts}
\usepackage{amssymb}

\def\endproof{\vrule height6pt width6pt depth0pt}

\begin{document}

\title{
Revisiting factorability and indeterminism}

\author{David Rodr{\'\i}guez}
\affiliation{Departamento de F{\'\i}sica Aplicada III, Universidad de
Sevilla, E-41092 Sevilla, Spain}
\email{drodriguez@us.es}

\date{\today}


\begin{abstract}

Perhaps it is not completely superfluous to remind that Clauser-Horne factorability,
introduced in \cite{CH74}, is only necessary when $\lambda$, the hidden variable
(HV), is \emph{sufficiently} deterministic:
for $\{M_i\}$ a set of possible measurements (isolated or not by space-like intervals)
on a given system, the most general sufficient condition for factorability
on $\lambda$ is obtained by finding a set of expressions $M_i=M_i(\lambda,\xi_i)$,
with $\{\xi_i\}$ a set of HV's, all independent from one another and from $\lambda$.
Otherwise, factorability can be recovered on $\gamma = \lambda \oplus \mu$,
with $\mu$ another additional HV, so that now $M_i=M_i(\gamma,\xi_i)$:
conceptually, this is always possible; experimentally, it may not:
$\mu$ may be unaccessible or even its existence unknown (and so, too, from the
point of view of a phenomenological theory).
Results here may help clarify our recent post in \cite{DR_WignerPDC}.

\end{abstract}



\maketitle

In relation to Bell inequalities, and maybe caused by the lack of a common
perspective, factorability and indeterminism are sometimes a subject of
prejudiced argumentation (at least at the informal level; that is my
experience); let us for that reason revisit those two concepts here.
We will try to settle a simple, completely abstract approach; not necessary
orthodox, we must warn.

\vspace{0.2cm}

\noindent\textbf{\emph{Definitions:}} \vspace{0.2cm}

We will say a measurement $M$ upon a certain physical system, with $k$ possible
outcomes $m_k$, is \emph{deterministic} on a hidden variable (HV) $\lambda$
(summarizing the state of that system), if (and only if)
\begin{eqnarray}
P(M = m_k |\lambda) \in \{0,1\}, \ \forall k, \lambda, \label{det}
\end{eqnarray}
which allows us to write 
\begin{eqnarray}
M \equiv M(\lambda),
\end{eqnarray}
and \emph{indeterministic} iff, for some $\lambda$, some $k^{\prime}$,
\begin{eqnarray}
P(M = m_{k^{\prime}} |\lambda) \neq \{0,1\}, \label{indet}
\end{eqnarray}
i.e., at least for some (at least two) of the results for at least
one (physically meaningful) value of $\lambda$.

\vspace{0.2cm}

Let also ${\cal M} = \{M_i\}$ be a set of possible measurements, each
with a set $\{m_{i,k}\}$ of possible outcomes, not necessarily isolated
from each other by a space-like interval.

\vspace{0.2cm}\noindent

Now, indeterminism can be turn into determinism, i.e., (\ref{indet})
can into (\ref{det}), by defining a new hidden variable $\mu$, so that
now, with $\gamma \equiv \lambda \oplus \mu$:
\begin{eqnarray}
P(M_i = m_{i,k} |\gamma) \in \{0,1\},
\ \forall i, \forall k,\gamma, \label{indet_ng}
\end{eqnarray}
which means we can write, for any of the $M_i$'s,
\begin{eqnarray}
M_i \equiv M_i(\lambda,\mu),
\end{eqnarray}
a proof that such a new hidden variable $\mu$ can always be found (or built)
given in \cite{proof_1}.

\vspace{0.2cm}\noindent

So far, then, our \emph{determinism} and \emph{indeterminism} are
conceptually equivalent, though of course they may correspond to different
physical situations, depending for instance on whether $\gamma$ is
experimentally accesible or not.
Nevertheless, for us there is still another natural step to take, introducing
the following distinction: we will say indeterminism is

\vspace{0.2cm}\noindent
(a) \emph{$\lambda$-factorizable}, iff
we can find a set $\{\xi_i\}$ of random variables, \emph{independent from
each other and from $\lambda$ too}, such that
\begin{eqnarray}
\mu = \bigoplus_{i} \xi_i, \label{indet_g_1}
\end{eqnarray}
and (\ref{det}) holds again for each $M_i$ on
$\gamma_i \equiv \lambda \oplus \xi_i$:
\begin{eqnarray}
P(M_i = m_{i,k} |\gamma_i) \in \{0,1\},
\ \forall i,\ \forall k, \gamma_i,
\label{indet_g_2}
\end{eqnarray}
this last expression meaning of course that we can write,
again for any of the $M_i$'s,
\begin{eqnarray}
M_i \equiv M_i(\lambda,\xi_i).
\end{eqnarray}

\vspace{0.2cm}\noindent
(b) \emph{non $\lambda$-factorizable}, iff
(\ref{indet_g_2}) is not possible for any set of statistically independent $\xi_i$'s.

\vspace{0.2cm}

Now let us, for simplicity, restrict our reasonings to $A,B \in {\cal M}$, with
two possible outcomes, $A,B \in \{+1,-1\}$, all without loss of generality.
We have seen that, as the more general formulation, we can always write something
like $A = A(\lambda,\xi_A)$, $B = B(\lambda,\xi_B)$.

\vspace{0.2cm}

\noindent\textbf{\emph{Lemma:}}

\vspace{0.2cm}\noindent
(i) \emph{If $A$ and $B$ are deterministic on $\lambda$},
i.e., (\ref{det}) holds for $A$ and $B$, \emph{then}
they are also $\lambda$-factorizable, i.e.,
\begin{eqnarray}
P(A=a,B=b|\lambda) = P(A=a|\lambda) \cdot P(B=b|\lambda), \label{fac}
\end{eqnarray}
for any $a,b \in \{+1,-1\}$.
Eq. (\ref{fac}) is nothing but the so-called Clauser-Horne factorability
condition \cite{CH74}.

\vspace{0.2cm}\noindent
(ii) \emph{If $A$ and $B$ are indeterministic on $\lambda$}, i.e.,
if (\ref{det}) does not hold for $\lambda$, \emph{then}:
for some $\mu$ (always possible to find \cite{proof_1}) such that now
(\ref{indet_ng}) holds for $\gamma \equiv \lambda \oplus \mu$,
$A,B$ are $\gamma$-factorizable,
\begin{eqnarray}
P(A=a,B=b|\gamma) = P(A=a|\gamma) \cdot P(B=b|\gamma), \label{fac_gamma}
\end{eqnarray}
i.e., (\ref{fac}) holds for $\gamma$, but this time not necessarily for
$\lambda$.

\vspace{0.2cm}\noindent
(iii) Let (\ref{indet_g_2}) hold for $A,B$, on $\lambda, \xi_A,\xi_B$:
if $\lambda, \xi_A,\xi_B$ are statistically independent,
(hence, $A$ and $B$ are what we have called $\lambda$-factorizable),
then (\ref{fac}) holds for $\lambda$, not necessarily on the contrary.

\vspace{0.2cm}

\noindent\textbf{\emph{Proof:}}

\vspace{0.2cm}\noindent
(i) When (\ref{det}) holds, for any $\lambda$ and any
$a,b \in \{+1,-1\}$, $P(A=a|\lambda), P(B=b|\lambda) \in \{0,1\}$,
from where we can, trivially, get to (\ref{fac}).
\hfill\endproof

\vspace{0.2cm}\noindent
(ii) It is also trivial that, if (\ref{indet_ng}) holds, (\ref{fac}) can be
recovered for $\gamma$.
That the same is not necessary for $\lambda$ can be seen with the following
counterexample:
suppose, for instance, that for $\lambda=\lambda_0$, either $A=B=1$ or
$A=B=-1$ with equal probability. It is easy to see that
\begin{eqnarray}
P(A=B=1|\lambda_0) \neq   P(A=1|\lambda_0) \cdot P(B=1|\lambda_0),
\label{counterex}
\end{eqnarray}
numerically: $\tfrac{1}{2} \neq \tfrac{1}{4}$.
\hfill\endproof

\vspace{0.2cm}\noindent
(iii) We have, from independence of $\lambda,\xi_A,\xi_B$, and
working with probability densities $\rho$'s:
$\rho_{\lambda}(\lambda,\xi_A,\xi_B) =
\rho_{\lambda}(\lambda) \cdot \rho_{A}(\xi_A) \cdot \rho_{B}(\xi_B)$,
which we can use to write
\begin{eqnarray}
&& P(A=a,B=b|\lambda) = \int P(A=a,B=b|\lambda,\xi_A,\xi_B) \nonumber\\
&&\quad\quad\quad\quad\quad\quad\quad\quad\quad\quad\quad
\times\ \rho_{A}(\xi_A) \cdot \rho_{B}(\xi_B) \ d\xi_A d\xi_B.
\nonumber\\
\end{eqnarray}
[Those conditioned probabilities should be defined also as
densities but for simplicity we leave that aside.]\\
Using now the fact that we can recover (\ref{indet_ng}) for $A$ ($B$) on
$\gamma_A = \lambda \oplus \xi_A$ ($\gamma_B = \lambda \oplus \xi_B$),
\begin{eqnarray}
&& P(A=a,B=b|\lambda)
\nonumber\\
&&\quad = \int
P(A=a|\lambda,\xi_A) \cdot P(B=b|\lambda,\xi_B) \nonumber\\
&&\quad\quad\quad\quad\quad\quad\quad\quad\quad\quad\quad
\times\ \rho_{A}(\xi_A) \cdot \rho_{B}(\xi_B)\ d\xi_A d\xi_B
\nonumber\\
&&\quad =
\int P(A=a|\lambda,\xi_A) \cdot \rho_{A}(\xi_A) \ d\xi_A \nonumber\\
&&\quad\quad\quad\quad\quad\quad\quad\quad
\times\ \int P(B=b|\lambda,\xi_B) \cdot \rho_{B}(\xi_B) \ d\xi_B
\nonumber\\
\nonumber\\
&&\quad = P(A=a|\lambda) \cdot P(B=b|\lambda).
\end{eqnarray}
On the other hand, let $\lambda, \xi_A,\xi_B$ be not statistically independent:
we can set for instance, as a particular case, $\xi_i \equiv \mu$, $\forall i$,
therefore reducing our case to that of (\ref{indet_ng}).
Once this is done, our previous counterexample in (ii) is also valid to show that
factorability is not necessary for $\lambda$ here.
\hfill\endproof

\vspace{0.1cm}

\noindent\textbf{\emph{Conclusions:}} 
In a bipartite (multipartite) Bell experiment, assuming information
is not degraded on its way from the source to the measurement
devices, $\xi_A,\xi_B$ ($\xi_i$) can be interpreted as the state of
devices $A,B$ (device $i$-th) and their surrounding, their independency
guaranteed by a space-like separation between observers.
\emph{
Given a theory that predicts results for the set ${\cal M}$ of measurements,
${\cal M}$ will be necessarily $\lambda$-factorizable only whenever all
relevant physical variables are actually included in the vector of hidden
variables $\lambda$.}

\vspace{0.1cm}

\noindent\emph{Acknowledgements:}
I thank R. Risco-Delgado for comments and encouragement.
We also acknowledge all the previous, surely in many cases more exhaustive,
work on the subject, in particular that of
Selleri and Tarozzi \cite{ST80,ST81}, as well as that of Risco-Delgado
himself \cite{Risco02}.




\begin{thebibliography}{99}
\bibitem{proof_1}
A possible (not unique) procedure to build $\mu$ is this:
for each $M_i$, we define a new random variable $\sigma_i$ and assign
values for each pair $(\lambda,m_{i,k})$:
$\sigma_i \equiv \sigma_i(\lambda,m_{i,k})$,
and now simply do
\begin{eqnarray}
\mu \equiv \bigoplus_{i} \sigma_i.
\end{eqnarray}
As built, $\sigma_i$'s are not necessarily independent from one another,
nor are they necessarily independent from $\lambda$.
\bibitem{DR_WignerPDC}
D. Rodr{\'i}guez,
``Wigner-PDC description of photon entanglement can still be made
completely local''.
Arxiv.
\end{thebibliography}
\end{document}